\documentclass[twocolumn,aps,reprint,noshowpacs,superscriptaddress]{revtex4}
\usepackage{graphicx}
\usepackage{bm}
\usepackage{amssymb}
\usepackage{amsmath}
\usepackage{amsthm}
\usepackage[colorlinks=true,linkcolor=blue,urlcolor=blue,citecolor=blue,pdfauthor={ },pdftitle={ },pdfsubject={ },pdfkeywords={ }]{hyperref}
\usepackage{array}

\graphicspath{{Figures/}}

\begin{document}
\title{Assessing three closed-loop learning algorithms by searching for \\ high-quality quantum control pulses}
\author{Xiao-dong Yang}
\affiliation{Hefei National Laboratory for Physical Sciences at the Microscale and Department of Modern Physics, University of Science and Technology of China, Hefei 230026, China}
\affiliation{CAS Key Laboratory of Microscale Magnetic Resonance and Department of Modern Physics, University of Science and Technology of China, Hefei, Anhui 230026, China}
\affiliation{Department of Chemistry, Princeton University, Princeton, New Jersey 08544, USA}

\author{Christian~Arenz}
\affiliation{Department of Chemistry, Princeton University, Princeton, New Jersey 08544, USA}

\author{Istvan~Pelczer}
\affiliation{Department of Chemistry, Princeton University, Princeton, New Jersey 08544, USA}

\author{Qi-Ming~Chen}
\affiliation{Department of Chemistry, Princeton University, Princeton, New Jersey 08544, USA}

\author{Re-Bing~Wu}
\email{rbwu@tsinghua.edu.cn}
\affiliation{Department of Automation, Tsinghua University \emph{\&} Center for Quantum Information Science and Technology, BNRist, Beijing 100084, China}

\author{Xin-hua Peng}
\email{xhpeng@ustc.edu.cn}
\affiliation{Hefei National Laboratory for Physical Sciences at the Microscale and Department of Modern Physics, University of Science and Technology of China, Hefei 230026, China}
\affiliation{CAS Key Laboratory of Microscale Magnetic Resonance and Department of Modern Physics, University of Science and Technology of China, Hefei, Anhui 230026, China}
\affiliation{Synergetic Innovation Centre of Quantum Information $\&$ Quantum Physics, University of Science and Technology of China, Hefei, Anhui 230026, China}

\author{Herschel~Rabitz}
\email{hrabitz@princeton.edu}
\affiliation{Department of Chemistry, Princeton University, Princeton, New Jersey 08544, USA}
\date{\today}

\begin{abstract}
Designing a high-quality control is crucial for reliable quantum computation. Among the existing approaches, closed-loop leaning control is an effective choice. Its efficiency depends on the learning algorithm employed, thus deserving algorithmic comparisons for its practical applications. Here we assess three representative learning algorithms, including GRadient Ascent Pulse Engineering (GRAPE), improved Nelder-Mead (NMplus), and Differential Evolution (DE), by searching for high-quality control pulses to prepare the Bell state. We first implement each algorithm experimentally in a nuclear magnetic resonance system and then conduct a numerical study considering the impact of some possible significant experimental uncertainties. The experiments report the successful preparation of the high-fidelity target state by the three algorithms, while NMplus converges fastest, and these results coincide with the numerical simulations when potential uncertainties are negligible. However, under certain significant uncertainties, these algorithms possess distinct performance with respect to their resulting precision and efficiency, and DE shows the best robustness. This study provides insight to aid in the practical application of different closed-loop learning algorithms in realistic physical scenarios.
\end{abstract}

\maketitle

\renewcommand{\thesubsection}{\arabic{subsection}}

\section{Introduction}
 Quantum control has drawn much attention in many areas \cite{goswami2003,bardeen1997,assion1998,brixner2004,silberberg2009}, such as quantum chemistry and quantum information \cite{nielsen2010quantum,brif2010control}. The state-of-the-art control design strategies are often based on modeling the system Hamiltonian and solving the dynamical evolution equations numerically. In practice, these efforts entail searching for an optimal control with iterative algorithms, including gradient-based \cite{khaneja2005optimal,riaz2019optimal} or gradient-free types \cite{zfk2001,zahedinejad2015high,ZG16,YL19}. However, a variety of uncertainties \cite{ferrie2015robust} exist in realistic experiments which can induce  unforeseen errors and distortions of the control design, thereby possibly preventing the realization of high-quality quantum control based on theoretical design for laboratory implementations.
 
\begin{figure}
  \centering
  \includegraphics[width=0.45\textwidth,height=0.25\textwidth]{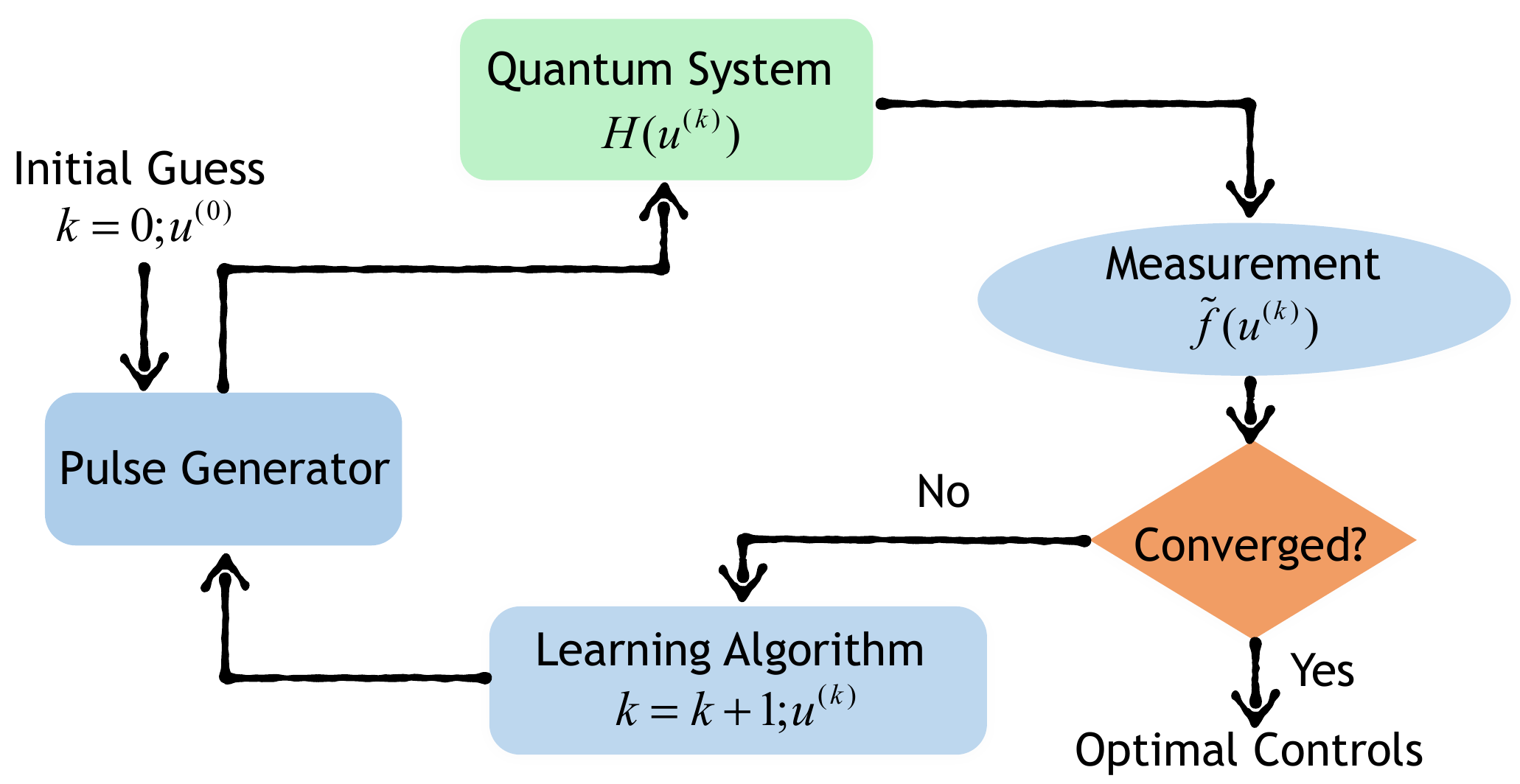}\\
  \caption{Schematic framework of closed-loop learning control, which entails the following steps: (1) Either randomly chosen or carefully designed initial controls $\bm{u}^{(0)}$ are applied to the quantum system. (2) The quantum system will evolve under a specific Hamiltonian $H({\bm{u}^{(k)}})$ on the $k$-th cycle around the loop. A suitable measurement is performed to specify the performance function $\widetilde {f}({\bm{u}^{(k)}})$. (3) The performance function obtained from the measurement is fed back to the learning algorithm to design new controls through specific updating laws. (4) These procedures are iteratively performed until the performance stopping criterion is met.}\label{Framework}
\end{figure}

 In order to address these issues in quantum control, several methods from classical control can be adapted \cite{chen2013closed}, such as robust control \cite{dong2015sampling,dong2016learning,egger2013optimized} and feedback control \cite{zhang2017quantum}. The present paper exploits closed-loop learning control (also called adaptive feedback control) pioneered by Judson and Rabitz \cite{judson1992teaching}, which is an effective method to defend against various uncertainties in practical experiments. The basic framework of closed-loop learning control is shown in Fig. \ref{Framework}. In the laboratory, a control performance function is used to guide the iterative optimization process. The model-free character of learning control in principle can ameliorate various uncertainties, imperfections and environment factors which are automatically considered in the process shown in Fig. \ref{Framework}. 
 
 In the case of quantum information science, the high-precision demand of the tasks present a challenge. Recent advances of applying closed-loop learning control in these areas include optimizing feasible quantum circuits by the algorithm MELVIN \cite{krenn2016automated}, finding system eigenstates with the help of variational algorithms \cite{peruzzo2014,wecker2015,mcclean2016} and creating quantum states and gates through gradient-based or evolutionary algorithms \cite{zahedinejad2015high,krenn2016automated,kelly2014optimal,egger2014adaptive,li2017hybrid,feng2018closed,CYA20}. The nature of the learning algorithm is a key component in determining the efficiency of the closed-loop optimization process. Thus, systematic comparisons of algorithmic options deserve attention for their practical applications and the identification of potential further improvements. 
 
It is not feasible here to exhaustively consider all possible learning algorithms. Rather, we explore three choices of algorithms with distinct characteristics.
To assess their performances,  we choose a typical quantum control problem of seeking to prepare a Bell state. We first experimentally apply the algorithms in a two-qubit nuclear magnetic resonance (NMR) system, followed by a numerical study including several possible significant experimental uncertainties. The paper begins by describing the high-quality pulse searching problem in Sec. \ref{Problem} along with introducing the three learning algorithms in Sec. \ref{AlgorithmIntro}. The experimental demonstrations follow in Sec. \ref{Exp}. Several significant uncertainties in the control process in diverse scenarios are considered numerically in Sec. \ref{Numerics}. The paper concludes with a discussion of the findings in Sec. \ref{Conclusion}.

 \section{Problem description} \label{Problem}
We begin by formulating the control problem for quantum state preparation examined in this paper. In order to express this problem generally, consider an $n$-qubit quantum system subject to transverse time-varying magnetic control fields ${\mathbf u}(t) = ({u_x^j}(t),{u_y^j}(t)):t \in [0,T]$, where $j=1,...,n$. The problem can be modelled as $H = {H_S} + \sum\nolimits_{j = 1}^n {{2\pi}({u_x^j}(t)\sigma _x^j + } {u_y^j}(t)\sigma _y^j)$, where $H_S$ represents the system internal Hamiltonian and $\sigma _x^j, \sigma _y^j$ denote the Pauli spin operators acting on the $j$-th spin. For the system starting out in a specific quantum state $\rho_0$, the control that operates over a finite time period $T$ will generate an evolution operator $U(T)$ which drives the system to some final state ${\rho _f} = U(T){\rho _0}{U^\dag }(T)$. We need a performance function to judge the effectiveness of the controls, such as the distance between $\rho_f$ and the target state $\rho_t$. If we consider that both $\rho_0$ and $\rho_t$ are pure states, then the state fidelity $F({\rho _f},{\rho _t})= \text{Tr}({\rho _f}{\rho _t})$ is a commonly employed performance function \cite{nielsen2010quantum}. Thus, the state preparation task can be formulated as
\begin{eqnarray}\label{problem}
&\max  &F({\rho _f},{\rho _t}) = \text{Tr}(U(T){\rho _0}U{(T)^\dag }{\rho _t}),\\ \nonumber
& \text{s.t.}  &\dot U(t) =  - i \left( {H_S} + \sum\limits_{j = 1}^n {2\pi} [ {u_x^j}(t)\sigma_x^j + {u_y^j}(t)\sigma _y^j] \right) U(t), 
\end{eqnarray}
where $U(0) = {I^{ \otimes n}}$ with $I$ being the $2 \times 2$ identity matrix. 

To optimize the controls in the present paper, we first divide the full evolution time $T$ into $M$ equal slices. Thus the time evolution operator for the $m$th interval $\Delta t=T/M$ can be expressed as
\begin{equation}
{U_m} = \exp \left\{  - i  \Delta t \left[ {H_S} +  \sum\limits_{j = 1}^n {2\pi} ({u_x^j}[m]\sigma _x^j + {u_y^j}[m]\sigma _y^j) \right] \right\}.
\end{equation}
The total evolution operator at time $T$ is then given by $U(T) = \prod \nolimits _{m=1}^M {U_m}$. Iterative learning algorithms are then used to optimize over the piecewise constant control field amplitudes $\bm{u}=(u_x^j[m],u_y^j[m])$, where $j=1,...,n,m=1, ...,M$, and the vector $\bm{u}$ contains $p=2nM$ elements. Thus the optimization goal is to maximize the fidelity $F(\rho_f,\rho_t) \equiv f(\bm{u})$ over $\bm{u}$. When the fidelity is measured in the laboratory, we will use $\widetilde f(\bm u)$ to distinguish it from the fidelity determined in the numerical simulations in Sec. \ref{Numerics}.

\section{Three illustrative learning algorithms} \label{AlgorithmIntro}
Many learning algorithms could be adapted to solve the above optimization problem in a closed-loop fashion described in Fig. \ref{Framework}. Though it is not feasible to exhaustively assess all possible learning algorithms, we explore three representative learning algorithms with distinct characteristics. The experimental and numerical studies in Secs. \ref{Exp} and \ref{Numerics}, respectively, give complementary insights into their performance. This type of assessment will necessarily be an ongoing effort in future studies by the community, including different applications and algorithms. Only then will a full picture of algorithmic choices become clear. The present work takes a significant step in that direction with the distinct algorithmic choices presented below.

\subsection{GRAPE}
GRadient Ascent Pulse Engineering (GRAPE) \cite{khaneja2005optimal} is an efficient optimal control algorithm. Though it was originally presented for NMR pulse optimization, it has been successfully adapted in other platforms \cite{haffner2008quantum,egger2014adaptive,Anderson2015accutate}. GRAPE starts either from some random or carefully designed initial controls $\bm u^{(0)}$, followed by iteratively updating the controls along the gradient ascent direction with an appropriate step length $\lambda^{(k)}$, i.e., $\bm u^{(k+1)} =\bm u^{(k)}+\lambda^{(k)} \partial f(\bm u^{(k)})/ \partial \bm u^{(k)}$. The algorithm terminates when the performance function stopping criterion is reached.

GRAPE is often used as an open-loop technique to design optimal controls for laboratory implementation, and it can provide reliable controls when good knowledge of the system is available. However, a variety of inevitable laboratory uncertainties may diminish the performance of the open-loop procedure. Consequently, GRAPE can be adapted to closed-loop operation in Fig. \ref{Framework} to address this issue. Besides measuring the gradient by finite-difference methods or statistical estimation strategies \cite{roslund2009gradient}, recently it was suggested to measure the gradient by inserting local rotations \cite{li2017hybrid}, which we will use in our study. The detailed procedure employed can be found in the Appendix.
  
\subsection{NMplus}
 The Nelder-Mead (NM) simplex algorithm \cite{nelder1965simplex} is a multidimensional direct search algorithm without use of the gradient. It starts from an initialized working simplex which consists of many vertices $\{ \bm u_1^{(0)}, \bm u_2^{(0)}, \cdots ,\bm u_{p + 1}^{(0)} \}$, where $p$ is the number of controls. Each vertex represents a vector of control parameters to be optimized. In each iteration, the worst performing vertex is replaced by a better one according to some specific geometric transformations, including reflection, expansion, contraction, and shrinkage over the simplex. These geometric transformations may be expressed by appropriate linear operators. The simplex moves towards the optimal solution direction iteratively until the performance function stopping criterion is met. Given the simplicity of the algorithm, it has been widely used in many quantum control experiments \cite{kelly2014optimal,egger2014adaptive,frank2017autonomous}. 
 
  However, the convergence speed of traditional NM is often very slow, which means that it may fail to achieve high-quality control performance over acceptable laboratory time. Fortunately, an accelerated strategy has been presented for updating the simplex, which can substantially improve the algorithmic efficiency \cite{pham2011comparative}.  In addition, we use the regular form of the initial simplex \cite{spendley1962sequential}. The improvements to the NM algorithm are referred to as NMplus, with further details found in the Appendix.
 
\subsection{Differential evolution}
Differential evolution (DE) \cite{das2011differential} is a gradient-free method within the evolutionary algorithm family. Like other members in this family, it is inspired by Darwinian principles and shares similar elements and procedures. More specifically, the parameterized controls of an optimization process are encoded into each individual expressed as a vector, and these parameters are usually randomly chosen to form the initial population $\{ \bm u_1^{(0)},\bm u_2^{(0)},...,\bm{u}_{P_n}^{(0)}\}$, where $P_n$ is the population size. With the algorithmic parameters being either fixed or adjusted adaptively, new individuals are generated through mutation and crossover operations, and then they are compared with the performance of the current individuals to decide on the survivors for the next generation through application of a selection strategy. These linear operations (mutation, crossover, selection) are applied to the entire population in the $k$th iteration. This procedure operates in the closed loop of Fig. \ref{Framework} until the performance function stopping criterion is met. DE has many favorable characteristics among the members of the evolutionary algorithm family. DE has also been implemented in many practical quantum control problems \cite{zfk2001,zahedinejad2015high,ZG16}. In the present study we apply an adapted efficient version of DE (DE/rand/2). The algorithmic details can be found in the Appendix.

\section{Experimental study}\label{Exp}
\begin{figure*}[htbp!]
  \centering
\includegraphics[width=0.95\textwidth,height=0.45\textwidth]{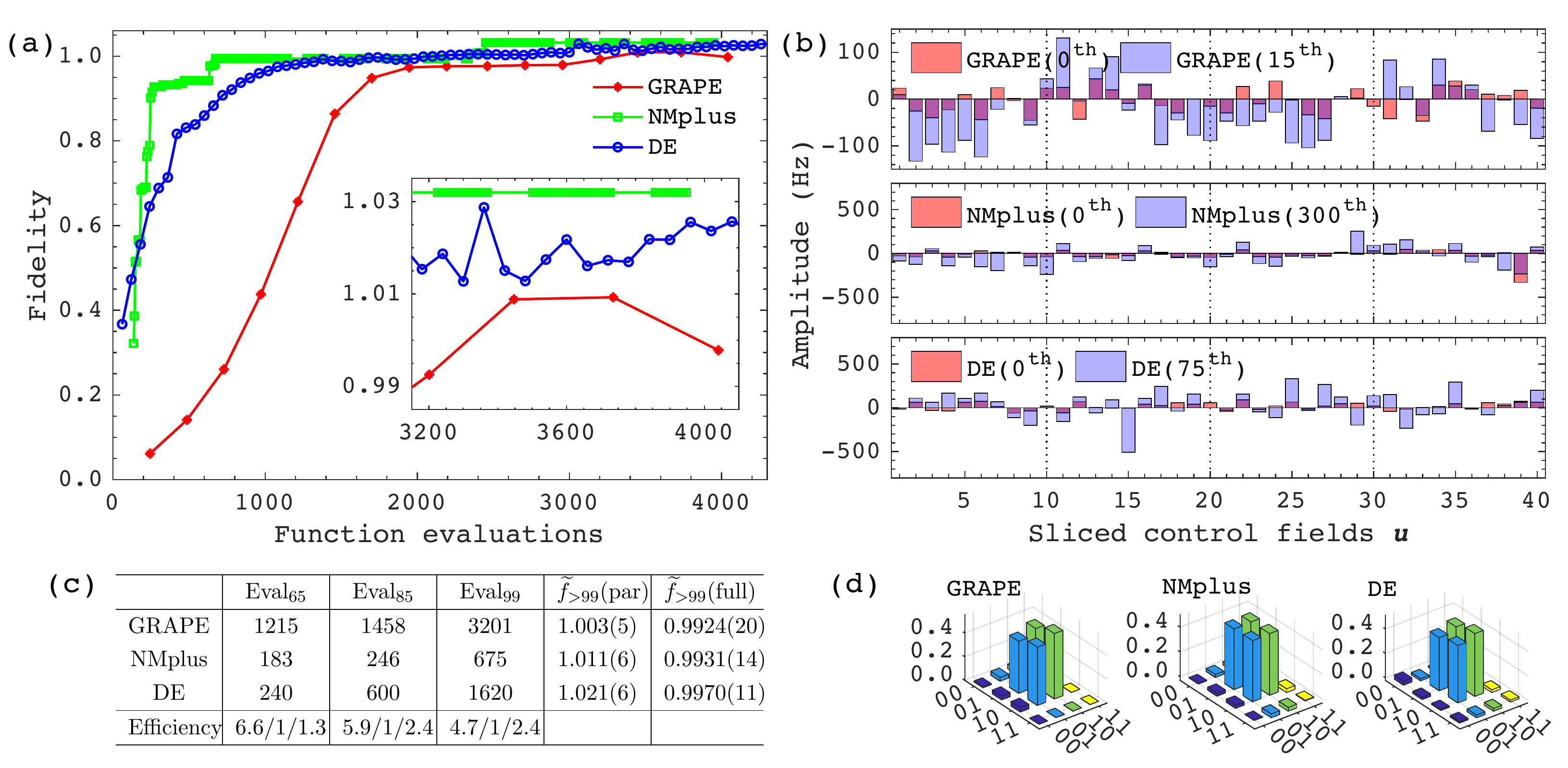}
  \caption{(Color online) Experimental closed-loop learning results for preparing the Bell state. The control pulses are all divided into $M=10$ slices, the total time period is {$T=5$ ms}, and the total number of the control parameters is $p=40$. The initial controls are all randomly chosen in the range $[-50,50]$ {Hz}. The maximum iteration number is set as $15$ for GRAPE, $300$ for NMplus, and $75$ for DE, respectively. 
  The algorithm parameters described in the Appendix are set as follows: (i) GRAPE. $\lambda^{(0)}=2*10^4,\lambda^{(l+1)}=0.5\lambda^{(l)},l\leq 40$; (ii) NMplus. $\alpha=3, \beta=1/3, \gamma=2, \delta=1/3$;  (iii) DE. $R=0.6,C_r=0.95,P_n=10$. (a) The entire experimental learning process, and the subplot enlarges the detailed performance approaching convergence. (b) Plots of the initial and final searched control fields $\mathbf{u}=(u_x^1[1],...,u_x^1[10],u_y^1[1],...,u_y^1[10],u_x^2[1],...,u_x^2[10],u_y^2[1],...,u_y^2[10])$ in the time sliced forms.
   (c) The function evaluations when the measured fidelity (partial state tomography) reaches around 0.65, 0.85, and 0.99, respectively. The experimental efficiency factor, defined by the ratio of function evaluations of each algorithm over function evaluations of NMplus, is also given. The final state fidelity with uncertainties (determined by applying the final controls to test the fidelity of the prepared Bell state ten times) obtained by partial and full state tomography are also listed. (d) Corresponding full state tomography results of the searched final state. 
}\label{ExpLoop}
\end{figure*}

\subsection{Experimental setup and procedure}
Our experiments for assessing these algorithms are conducted using chloroform ($\rm CHCl_3$) as a sample on a Bruker Avance III 400-MHz spectrometer at room temperature. The sample consists of two nuclear spins $^{13}\text{C}$ and $^{1}\text{H}$, respectively labeled as 1 and 2. The internal Hamiltonian in the rotating frame can be simply expressed as ${H_S} =\pi J_{12}\sigma_z^1\sigma_z^2/2$ \cite{vandersypen2005nmr}, where $J_{12}=214.5~\rm{Hz}$ is the scalar coupling strength between the two spins.

We first initialize the system into a pseudopure state ${\rho_{\mathrm{pps}}} = \frac{{1 - \varepsilon }}{4}I + \varepsilon \left| {00} \rangle \langle {00} \right|$ by the line-selective method \cite{PZX01}, where $\varepsilon \approx 10^{-5}$ represents the thermal polarization of the two-qubit system. As the identity matrix has no physical observable effects, the initial state can be treated as $\rho_0=\left| {00} \rangle\langle {00} \right|$. Full state tomography \cite{LJS02} verifies that the experiment realized  $\rho_0$ with a fidelity of $0.9976$. From this initial state, the three learning algorithms are separately applied to search for optimal pulses that iteratively transform $\rho_0$ as close as possible to the target Bell state $\rho_t=\left| {{\psi _t}} \rangle \langle {{\psi _t}} \right|$ with $\left| {{\psi _t}} \right\rangle  = (\left| {10} \right\rangle  + \left| {01} \right\rangle )/\sqrt 2 $ . To compare these algorithms fairly, the control fields are all divided into $M=10$ slices with a total time evolution $T=5~\text{ms}$. 

Experimentally evaluating the prepared state is always a resource-consuming task. A suitable measurement strategy should consider the properties of the target state and the corresponding experimental resources needed \cite{gross2010quantum, knill2008randomized,granade2016practical}. Here we choose the partial state tomography method to measure the iteratively evolving state fidelity. Specifically, the target two-qubit Bell state $\left| {{\psi _t}} \right\rangle  = (\left| {10} \right\rangle  + \left| {01} \right\rangle )/\sqrt 2 $ can be decomposed in the Pauli basis, i.e., ${\rho _t} = \left| {{\psi _t}} \rangle \langle {{\psi _t}} \right| = (I^{\otimes{2}} + {\sigma _x^1}{\sigma _x^2} + {\sigma _y^1}{\sigma _y^2} - {\sigma _z^1}{\sigma _z^2})/4$. This indicates that measurements using the basis operators $\sigma_x^1\sigma_x^2, \sigma_y^1\sigma_y^2$ and $\sigma_z^1\sigma_z^2$ are sufficient to estimate the state fidelity between the final state $\rho_f$ and the target $\rho_t$. Here we define each measurement on one of the above basis members as a single \textit{function evaluation}.

The algorithmic operational procedures are then briefly described as follows; the details can be found in the Appendix.

\textit{GRAPE}: (1) The initial control fields $\bm{u}^{(0)}=(u_x^j[m]^{(0)},u_y^j[m]^{(0)})$, where $ u_\alpha^j[m]^{(0)} \in [-50, 50]{ ~\text{Hz}}; \alpha=x,y; j=1,2; m=1,2,...,10$, are randomly generated on a classical computer.
(2) The initial controls are applied to the NMR processor, which transform $\rho_0$ to some quantum state, and its fidelity $\widetilde f({\bm{u}^{(0)}})$ is measured by the introduced partial state tomography method. Meanwhile, the gradients of the $m$th sliced control, i.e.,  $g_{\alpha}^j[m]^{(0)}=\partial \widetilde f(u_{\alpha}^j [m]^{(0)}) / \partial u_\alpha^j[m]^{(0)}$, are measured by inserting local rotations $\{R_\alpha ^j( \pm  \frac{\pi }{2})| \alpha=x,y, j=1,...,n\}$ (see details in Appendix).
(3) From $k=0$, new controls are updated by the law introduced in Sec. \ref{AlgorithmIntro} A, and the corresponding fidelity of the prepared state $\widetilde f({\bm{u}^{(k+1)}})$ is measured again. 
(4) The iteration number is set as $k\rightarrow k+1$, and this procedure is looped until the maximum iteration $15$ (sufficient to reach convergence) is attained. 
In this procedure, we need $3(4nM+1)=243$ function evaluations in each iteration, including the gradient measurements.

\textit{NMplus}: (1) A regular form of the initial simplex is formed which consists of $p+1=2nM+1=41$ vertices generated on a classical computer, and each vertex represents a vector of candidate control pulses, i.e., $\{ \bm u_1^{(0)}, \bm u_2^{(0)}, \cdots ,\bm u_{p+1}^{(0)} \}$; $\bm u_i^{(0)} = (u_{i1}^{(0)},...,u_{ip}^{(0)}); u_{ij}^{(0)} \in [-50,50] {~\text{Hz}},i=1,2,...,41,j=1,2,...,40$. 
(2) The initial candidate controls are applied to the NMR processor sequentially to generate candidate quantum states, and their fidelity $\{\widetilde f({\bm{u}_i^{(0)}})\}$ is measured by the partial state tomography method.
(3) From $k=0$, new controls are used to refresh the simplex by the updated law introduced in Sec. \ref{AlgorithmIntro} B, and the corresponding fidelity of the prepared quantum states $\{\widetilde f({\bm{u}_i^{(k+1)}})\}$ are measured again, where $i=1,2,...,41$. 
(4) The iteration number is set as $k\rightarrow k+1$, and this procedure is repeated until the maximum iteration $300$ (sufficient to reach convergence) is attained. In this procedure, the function evaluations in each iteration are $3w$, where $w \in (1,2,3,4,5)$ depending on specific algorithmic procedure.

\textit{DE}: (1) An initial population containing $P_n=10$ individuals $\{ \bm u_1^{(0)},\bm u_2^{(0)},...,\bm{u}_{P_n}^{(0)}\}$ is randomly generated on the classical computer, and each individual in this population represents the genes of the candidate control pulses, namely $\bm u_i^{(0)} = (u_{i1}^{(0)},...,u_{ij}^{(0)});u_{ij}^{(0)} \in [-50,50] {~\text{Hz}},i=1,2,...,10,j=1,2,...,10$. 
(2) These candidates are applied to the NMR processor to generate quantum states, and their fidelity $\{\widetilde f({\bm{u}_i^{(0)}})\}$ is measured by the partial state tomography method.
(3) From $k=0$, new controls are updated to refresh the population by the laws introduced in Sec. \ref{AlgorithmIntro} C, and the corresponding fidelity of the prepared quantum states $\{\widetilde f({\bm{u}_i^{(k+1)}})\}$ are measured again, where $i=1,2,...,10$.
(4) The iteration number is set as $k\rightarrow k+1$, and this procedure is looped until the maximum iteration $75$ (sufficient to reach convergence) is attained. In this procedure there are $6 P_n=60$ function evaluations in each iteration.

\subsection{Experimental results}
The experimental results of preparing the target Bell state by closed-loop learning are shown in Fig. \ref{ExpLoop}. In general, the results suggest that all of the three algorithms successfully prepare the target state with fidelities close to or even beyond $1$ from Fig. \ref{ExpLoop}(a); in this regard, note that the incomplete measurements by partial state tomography can slightly distort the true fidelity. Thus we utilize full state tomography to verify these results, as shown in Figs. \ref{ExpLoop}(c) and \ref{ExpLoop}(d), where we define the measured fidelity above 0.99 by partial state tomography and full state tomography as $\widetilde{f}_{>99}(\text{par})$ and $\widetilde{f}_{>99}(\text{full})$, respectively. The statistical errors of the fidelities are obtained by applying the final control ten times to calculate the standard deviation. The results reveal that the target state is prepared with fidelity above 0.99 by all three algorithms. Additionally, partial state tomography is shown to give reliable guidance  during optimization, rather than using resource-consuming full tomography. To compare the convergence speed of these algorithms, we list the required function evaluations when the fidelity approaches $0.65,0.85$, and $0.99$ (marked as $\text{Eval}_{65}$, $\text{Eval}_{85}$, and $\text{Eval}_{99}$, respectively) in Fig. \ref{ExpLoop}(c), from which we find NMplus can achieve up to six times faster convergence speed than the other two algorithms. This behavior is already evident in Fig. \ref{ExpLoop}(a). We also plot the initial control fields and the final discovered optimal control fields for each algorithm in Fig. \ref{ExpLoop}(b).

Such high fidelities were achieved due to the nature of closed-loop learning control described above and the isolated and well-controlled NMR system \cite{vandersypen2005nmr}. We now briefly discuss the experimental uncertainties that contribute to the learning procedure: 
(1) Measurement errors. They can be estimated by adding up (i) the classical instrument noise, which is characterized by the noise-to-signal ratio ($\sim 0.0004$ in our instrument); (ii) the random measurement error, which can be inferred by applying the same control sequence five times to obverse the variations of the measurement results, which is at a level of $\sim 0.0005$. In total, these errors induce a uncertainty of $\sim 0.0009$, which coincides with the full state tomography results.
(2) Initial state preparation errors, control imperfections, and decoherence. They can be estimated by applying an open-loop determined pulse to the NMR sample, then comparing the measured fidelity with its numerically computed value. As such, implementing an open-loop searched pulse of fidelity $f_{\text{sim}}=0.9965$, the experimentally measured fidelities for five repetitive times are $0.9930, 0.9935, 0.9926, 0.9938, 0.9933$ through full state tomography. The standard deviation of the measured value and the simulated value is calculated to be $\sigma_c=\sqrt{\sum_{i=1}^N(f_{\text{exp}}^i-f_{\text{sim}})^2/N}=0.0033$. Note that the above analyzed measurement errors are also involved in this deviation.
{While deducting the measurement errors at a level of $\sim 0.0009$, a deviation of 0.0024 still remains. However, Fig. \ref{ExpLoop}(c) shows that the measured deviation of full state tomography is much smaller (0.0020/0.0014/0.0011); this result reveals that in the closed-loop learning process, part of the errors coming from the above experimental uncertainty (2) are corrected.}

\section{Numerical study}\label{Numerics}
The numerical simulations are performed to bolster the experimental findings, as well as give some clues for circumstances beyond NMR where larger experimental imperfections and errors can exist. Each of the three algorithms is explored in this fashion.

\begin{figure*}[htbp!]
  \centering
\includegraphics[width=0.9\textwidth,height=0.43\textwidth]{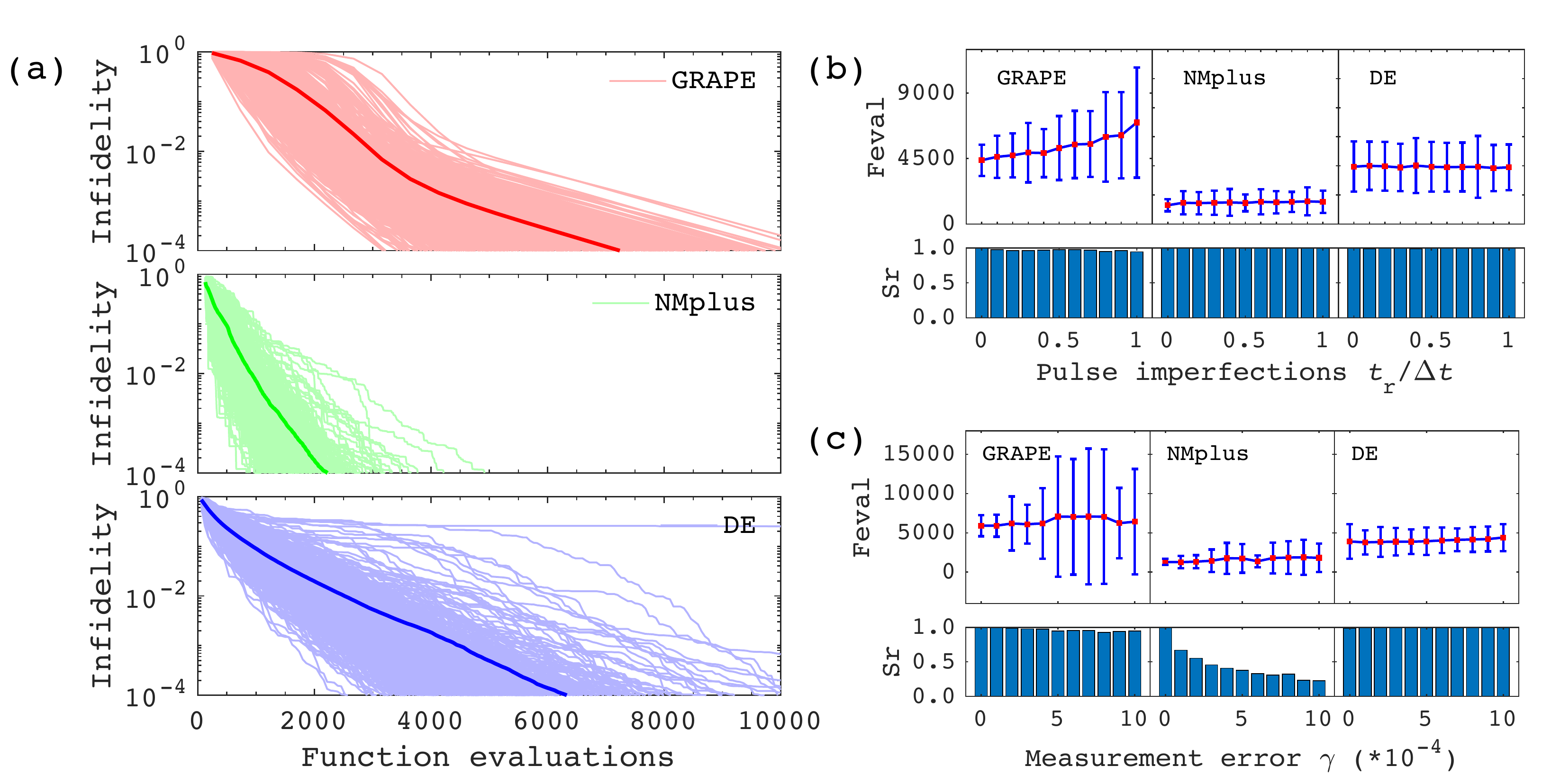}\\
  \caption{(Color online) Numerical search results with 500 runs of the three algorithms seeking optimal pulses to prepare the Bell state. 
  (a) State infidelity ($1-F$) vs function evaluations. The thick lines are the statistical average of the data over 500 runs in each case.  
  (b), (c) Function evaluations (abbreviation Feval) utilized when searching for the pulses to prepare the Bell state in the presence of pulse imperfections and measurement errors, respectively. Each algorithm stops when the state infidelity reaches 0.001 within $10^5$ function evaluations. Thus we exclude the failed runs and give the corresponding success rates (the rate of runs reaching infidelity 0.001 over 500 runs, abbreviation Sr) at the bottom. The red squares show the mean function evaluations, and the blue bars are the corresponding variance.}\label{Simulation}
\end{figure*}

\subsection{Direct run}
The experiments in Sec. \ref{Exp} show the results in a single run of each algorithm. Here we assume potential uncertainties in the control learning process are negligible and perform numerical simulations of the algorithms with each running 500 times to further assess their performance.
We show the state infidelity ($1-F$) with respect to the function evaluations in Fig. \ref{Simulation}(a). The thick lines are the statistical average of the corresponding infidelities at each number of function evaluations over 500 runs. The final infidelities of a few runs in DE that do not reach $10^{-4}$ are ignored. Also, small differences of the function evaluations in each run for NMplus (the function evaluations in each iteration are not constant, as stated in Sec. \ref{Exp} B) are ignored when calculating the statistical average. The above experimental results are consistent with the numerical simulations (are characterized by the statistical average results), which indicates that the numerical simulations are likely capable of giving reliable predictions of experimental applications in an NMR system beyond the Bell state goal. Furthermore, these results indicate that GRAPE and DE have similar convergence speeds in reaching low infidelity, while NMplus converges significantly faster for preparing the Bell state. It is also evident that the function evaluations corresponding to the infidelity $10^{-4}$ are distributed over a broader range for GRAPE and DE than that for NMplus. This result indicates that NMplus performs more consistently in this case.  

\subsection{Control imperfections}
The laboratory-generated control fields can suffer from a variety of technical limitations, which we collectively refer to as pulse imperfections. Such uncertainties could restrict the ability of the control to achieve expected quantum system performance, particularly beyond the NMR setting. To some degree, learning control can fight against such limitation, but the convergence speed of the learning algorithms may be influenced. Additionally, it is not possible to exhaustively know or assess the impact of all possible control imperfections. Here we consider the illustrative case of control distortion, which is described by a liner filter \cite{wu2018data}: $\bm v(t)=\mathcal{D}[\bm u(t)]=\int_0^{\infty}h(t-\tau) \bm u(t) d\tau	$, where $h(t)=\frac{1}{t_r}e^{-\frac{t}{t_r}},t \geq 0$ is the impulse function. The time constant $t_r$ characterizes the pulse distortion, where a large value of $t_r$ indicates significant distortion of the pulses. We vary $t_r/\Delta t$ from $0$ to $1$ with the stepsize $0.1$ to assess the performance of the three learning algorithms to this type of pulse distortion. As shown in Fig. \ref{Simulation}(b), for both NMplus and DE, increasing $t_r/\Delta t$ does not greatly influence the averaged function evaluations. However, for GRAPE, the averaged function evaluations increase as $t_r/\Delta t$ becomes larger. This behavior is easily understood, as the control imperfections will induce errors in estimating the gradient, thus influencing the convergence speed. NMplus and DE only need the fidelity information, which is not significantly influenced by these particular pulse imperfections. In addition, GRAPE and DE have relatively large variances, indicating unstable performance compared to NMplus.

\subsection{Measurement errors}
Due to the technical precision limitations and the statistical properties of the measurement strategies, the measured performance functions always deviate from their true values. Such measurement errors can be divided into two types: systematic and random. As an illustration, here we consider an additive random measurement error which obeys a normal distribution with variance $\gamma^2$ and mean zero, $f(\bm u)\rightarrow f(\bm u) + N(0,\gamma^2)$. In Fig. \ref{Simulation}(c), $\gamma$ is varied from $0$ to {$0.001$}, with the step size being {$0.0001$}, to compare the performance of the three algorithms. {We find that as the measurement error variance increases, the function evaluations that GRAPE needs to reach the target fidelity vary erratically, while the variation of the function evaluations for NMplus and DE are much smaller.} The reason for this behavior is that the gradient estimations are very sensitive to the measurement errors. Large measurement errors can cause {difficulty} in identifying the ``right'' climbing direction, thus leading to {distinct convergence rates} \cite{roslund2009gradient}. In addition, for large measurement errors, NMplus has very low success rates. This is due to the success of NMplus greatly relying on the directions of the geometric transformations of the simplex. In general, large measurement errors will produce false directions and possibly lead to failure upon searching for the target \cite{egger2014adaptive}. 

\section{Discussion and conclusion} \label{Conclusion}
We experimentally and numerically assessed GRAPE, NMplus, and DE by seeking optimal pulses for preparing a Bell state to high fidelity. The experiments reported finding high-quality pulses by all the three learning algorithms, and NMplus achieved convergence up to six times faster than the other two algorithms. The performance of the three algorithms in the numerical simulations was in agreement with that in the experiments. This outcome indicates that the overall uncertainties in NMR platform are small so that the numerical simulations likely can give reliable predictions on other targetable states or gates if the Hamiltonian is well known. However, for many other quantum systems outside of an NMR control setting, the uncertainties may significantly influence the control performance. Thus, as illustrations we considered two common uncertainties: pulse distortion and random measurement error. These effects were assessed in the Bell state preparation. GRAPE was found to be more sensitive to these two uncertainties, NMplus shows low success rate to deal with large random measurement errors, while DE is the most robust for both types of uncertainties.

Closed-loop learning is a promising method to achieve high-quality and efficient quantum control, which has attracted increasing attention. The numerical simulations show their different features under particular assumed forms of uncertainties, thus giving clues about which algorithm to apply in other settings besides NMR. However, a fully general conclusion cannot be drawn about algorithmic choices, as many factors can enter. Further investigations should consider a broad variety of control goals and even additional algorithmic choices in a closed-loop algorithmic setting.

\section*{ACKNOWLEDGEMENTS}
X.P. acknowledges the National Key Research and Development Program of China (Grant No. 2018YFA0306600), the National Natural Science Foundation of China (Grants No. 11661161018 and No. 11927811), and the Anhui Initiative in Quantum Information Technologies (Grant No. AHY050000), and C.A. acknowledges the US ARO (Grant No. W911NF-19-1-0382). R.W. acknowledges the support of the NSFC (Grants No. 61833010 and No. 61773232). H.R. acknowledges the US DOE (Grant No. DE-FG02-02ER15344), and C.A. acknowledges the US ARO (Grant No. W911NF-19-1-0382).

\bibliographystyle{apsrev4-2.bst}
\providecommand{\noopsort}[1]{}\providecommand{\singleletter}[1]{#1}%

\begin{appendix}
\renewcommand{\thesubsection}{\arabic{subsection}}
\renewcommand{\theequation}{A\arabic{subsection}}
\section*{APPENDIX}
In this Appendix, we present the details of the three closed-loop learning algorithms GRAPE, NMplus, and DE. We use $\bm{u}=(u_x^j[m],u_y^j[m])$ to represent the controls to be optimized, where $j=1,...,n$ is the qubit number, and $m=1,...,M$ is the number of the control slices. We then denote the $k$th iteration of the controls as $\bm{u}^{(k)}=(u_x^j[m]^{(k)},u_y^j[m]^{(k)})$. As the vector $\bm{u}$ contains $p=2nM$ elements, the controls can then be expressed in another equivalent form $\bm{u}^{(k)}=(u_1^{(k)},u_2^{(k)},...,u_p^{(k)})$. The corresponding measured performance function of $\bm{u}^{(k)}$ is expressed as $\widetilde f(\bm{u}^{(k)})$.

\subsection{GRAPE}
GRadient Ascent Pulse Engineering (GRAPE) starts from an initial guess $\bm u^{(0)}$, and new controls $\bm u^{(k+1)}$ are generated by a linear strategy 
\begin{equation}
u_{\alpha}^j [m]^{(k+1)}=u_{\alpha}^j[m]^{(k)}+\lambda^{(l)} g_{\alpha}^j[m]^{(k)}, 	
\end{equation}
where $g_{\alpha}^j[m]^{(k)}=\partial \widetilde f(u_{\alpha}^j [m]^{(k)}) / \partial u_\alpha^j[m]^{(k)}$ is the gradient of the measured fitness function, $\lambda^{(k)}$ is an appropriate step length updated by $\lambda^{(l+1)}=0.5\lambda^{(l)}$ in each iteration, and $\alpha=x,y$. The closed loop terminates when the performance stopping criterion is satisfied. As described in Ref. \cite{li2017hybrid}, we can obtain the gradient by 
\begin{equation}
g_{\alpha}^j[m] = \Delta t {[\text{Tr}(\rho _{{\alpha _ + }}^{jm}{\rho _t}) - \text{Tr}(\rho _{{\alpha _ - }}^{jm}{\rho _t})} ],	
\end{equation}
 where $\rho _{{\alpha _ \pm }}^{jm} = U_{m + 1}^MR_\alpha ^j( \pm \frac{\pi }{2})U_1^m{\rho _0}{(U_{m + 1}^MR_\alpha ^j( \pm \frac{\pi }{2})U_1^m)^\dag }$, and $U_{m_1}^{m_2}$ denotes ${U_{{m_2}}} \cdots {U_{{m_1} + 1}}{U_{{m_1}}}$. This means that by adding in the finite operation set of single-qubit rotations $\{R_\alpha ^j( \pm  \frac{\pi }{2})| \alpha=x,y, j=1,...,n\}$, one can measure the gradient information directly. In the main text, the experiments used this gradient measurement strategy.

\subsection{NMplus}
 The Nelder-Mead plus (NMplus) algorithm is operated as follows:
  
\textbf{Step 1}: set algorithm constants $\alpha, \gamma, \beta, \delta$, generate an initial simplex with vertices $\{ \bm u_1^{(0)}, \bm u_2^{(0)}, \cdots ,\bm u_{p + 1}^{(0)} \}$, where $\bm u_i^{(0)} = (u_{i1}^{(0)},...,u_{ip}^{(0)})$, and calculate their function values according to the measured performance functions $\widetilde f(\bm u_i^{(0)})$, i.e., $f_i^{(0)}=1-\widetilde f(\bm u_i^{(0)}), i=1,2,...,p+1$.

\textbf{Step 2}:  sort the vertices at the $k$-th iteration so that $f_1^{(k)}\le f_2^{(k)} \le  \cdots  \le f_{p + 1}^{(k)}$, where $f_i^{(k)}=1-\widetilde f(\bm u_i^{(k)}), i=1,2,...,p+1$.

\textbf{Step 3}: the $p+1$ vertices form a hyperplane with the approximate multiple equations $f_i^{(k)} = {a_0^{(k)}} + {a_1^{(k)}}{u_{i1}^{(k)}} + {a_2^{(k)}}{u_{i2}^{(k)}} + ... + {a_p^{(k)}}{u_{ip}^{(k)}},i = 1,2,...,p + 1$.   Calculate the approximate reflection direction by writing the above multiple equations in matrix form: 
\begin{eqnarray}
\begin{split}
&\qquad\qquad\quad \bm G=\bm X^{-1}\bm Y,	\\ 
&\bm G=[a_0^{(k)}~a_1^{(k)}~...~a_p^{(k)}]^T,  
 \bm Y=[f_1^{(k)}~f_2^{(k)}~...~f_{p+1}^{(k)}]^T, \\
&\bm X=(x_{i,j}^{(k)})_{(p+1) \times (p+1)}, x_{i,j+1}^{(k)}=u_{ij}^{(k)},x_{i,1}^{(k)}=1. 
\end{split}
\end{eqnarray} 
Calculate the reflection point, ${\bm u_r^{(k)}} = \bm u_1^{(k)} - \alpha* \bm G$, and evaluate the function value $f_r^{(k)}=1-\widetilde f(\bm u_r^{(k)})$.

\textbf{Step 4}:
\textbf{S4.1}: if $f_1^{(k)} \le f_r^{(k)} < f_p^{(k)}$, let $\bm u_{p+1}^{(k)}=\bm u_r^{(k)}, f_{p+1}^{(k)}=f_r^{(k)}$.

\textbf{S4.2}: if $f_r^{(k)} < f_1^{(k)}$, calculate the expansion point, ${\bm u_e^{(k)}} = \bm u_1^{(k)}  + \gamma(\bm u_r^{(k)} - \bm u_1^{(k)})$, evaluate the function value $f_e^{(k)}=1-\widetilde f(\bm u_e^{(k)})$. \\
(a) if $f_e^{(k)} < f_r^{(k)}$, let $\bm u_{p+1}^{(k)}=\bm u_e^{(k)}, f_{p+1}^{(k)}=f_e^{(k)}$.\\
(b) if $f_e^{(k)} > f_r^{(k)}$, let $\bm u_{p+1}^{(k)}=\bm u_r^{(k)}, f_{p+1}^{(k)}=f_r^{(k)}$.

\textbf{S4.3}: if $f_r^{(k)} \ge  f_{p}^{(k)}$:\\
(a) if $f_p^{(k)} \le f_r^{(k)} < f_{p+1}^{(k)}$, calculate the outside contraction point, ${\bm u_c^{(k)}} = \bm u_1^{(k)} + \beta (\bm u_r^{(k)} - {\bm u_1^{(k)}})$, evaluate the function value $f_c^{(k)} = 1-\widetilde f(\bm u_c^{(k)})$:\\
$(a_1)$ if $f_c^{(k)} \le f_r^{(k)}$, let $\bm u_{p+1}^{(k)}=\bm u_c^{(k)}, f_{p+1}^{(k)}=f_c^{(k)}$;\\
$(a_2)$ if $f_c^{(k)}  > f_r^{(k)}$, shrink the simplex, ${\bm u_i^{(k)}} = {\bm u_1^{(k)}} + \delta(\bm u_i^{(k)} -  {\bm u_1^{(k)}}),{f_i^{(k)}} =1-\widetilde f({\bm u_i^{(k)}}),i = 2,3, \cdots ,p + 1$;\\
(b) if $f_r^{(k)} \ge f_{p+1}^{(k)}$, calculate the inside contraction point, ${\bm u_c^{(k)}} = \bm u_1^{(k)} - \beta({\bm u_{r}^{(k)}} - {\bm u_1^{(k)}})$, evaluate the function value $f_c^{(k)} = 1-\widetilde f(\bm u_c^{(k)})$:\\
$(b_1)$ if $f_c^{(k)} \le f_r^{(k)}$, let $\bm u_{p+1}^{(k)}=\bm u_c^{(k)}, f_{p+1}=f_c^{(k)}$;\\
$(b_2)$ if $f_c^{(k)}  > f_r^{(k)}$, shrink the simplex, ${\bm u_i^{(k)}} = {\bm u_1^{(k)}} + \delta(\bm u_i^{(k)} -  {\bm u_1^{(k)}}),{f_i^{(k)}} =1-\widetilde f({\bm u_i^{(k)}}),i = 2,3, \cdots ,p + 1$.

\textbf{Step 5}: Check the performance stopping conditions, and if they are not satisfied, let $k=k+1$ and continue with step 2.

To further improve the algorithm efficiency, the normal form of the initial simplex \cite{spendley1962sequential} is slightly modified:
\begin{equation}
u_{ij}^{(0)}=\left\{
\begin{array}{*{10}{c}}
u_{1j}^{(0)}+ \frac{C_{ij}}{\sqrt p} (\sqrt{p+1}-1)  && {i\ne j+1 ~\&\&~ i > 1}\\
u_{1j}^{(0)}+ \frac{C_{ij}}{\sqrt p} (\sqrt{p+1}+p-1)  && {i=j+1 ~\&\&~ i > 1},
\end{array} \right.
\end{equation}
where $u_{1j}^{(0)}=0$ and $C_{ij}$ represents some random amplitude of the control fields. Together with the above quasigradient NM algorithm, we call it NMplus.

\subsection{DE}
The differential evolution (DE) algorithm used in this study is described as follows: 

\textbf{Step 1}: set the algorithm constants: scaling factor $R$, crossover rate $C_r$, chromosome length (the dimension of each individual, namely, the length of the controls to be optimized) $p$ and population size $P_n$. Generate an initial population $P= \{ \bm u_1^{(0)},\bm u_2^{(0)},...,\bm{u}_{P_n}^{(0)}\}$ randomly with $\bm u_i^{(0)} = (u_{i1}^{(0)},...,u_{ip}^{(0)})$ being the $i$th individual in current population. 

\textbf{Step 2}: From $i=1$ to $P_n$, do the following steps at iteration $k$: \\
(1) Mutation.  Generate a donor vector $\bm v_i^{(k)}=(v_{i1}^{(k)},...,v_{ip}^{(k)})$ through the following mutation rule: 
\begin{equation}
{\bm v_i^{(k)}} = {\bm u_{r_b^i}^{(k)}} + R({\bm u_{r_1^i}^{(k)}} - {\bm u_{r_2^i}^{(k)}}+{\bm u_{r_3^i}^{(k)}} - {\bm u_{r_4^i}^{(k)}}),	
\end{equation}
 where $r_1^i, r_2^i, r_3^i, r_4^i$ are randomly chosen mutually exclusive integers in the range $[0, P_n]$ and $r_b^i$ is the index of the best individual in the current population.\\
(2) Crossover. Generate a trial vector $\bm \mu_i^{(k)}=(\mu_{i1}^{(k)},...,\mu_{ip}^{(k)})$ by a binomial crossover strategy: if $rand_{i,j}[0,1] \le C_r$ or $j=j_{rand}$, let $\mu_{ij}=v_{ij}$, where $j_{rand} \in \{1,2,...,p\}$ is a randomly chosen index. Otherwise let $\mu_{ij}=u_{ij}$. \\
(3) Selection. Evaluate the former individual $\bm u_i^{(k)}$ and the trail vector $\bm \mu_i^{(k)}$, if $\widetilde f(\bm\mu_i^{(k)})\ge \widetilde f(\bm u_i^{(k)})$, let $\bm u_i^{(k)}=\bm \mu_i^{(k)}$, otherwise keep $\bm u_i^{(k)}$ unchanged. 

\textbf{Step 3}: Check the performance stopping criterion, and if it is not satisfied, let $k=k+1$ and continue with step 2.

\end{appendix}

\end{document}